\documentclass[pra,twocolumn,groupedaddress]{revtex4-1}
\usepackage[T1]{fontenc}
\usepackage[english]{babel}
\usepackage{siunitx}
\usepackage[utf8]{inputenc}
\usepackage{url}
\usepackage{dsfont}
\usepackage{bbm}
\usepackage{nicefrac}
\usepackage{setspace}
\usepackage[short]{optidef}
\usepackage{float}
\usepackage{comment}
\usepackage{dsfont}
\usepackage{amssymb}  
\usepackage[colorlinks=true, citecolor=blue,urlcolor=blue,linkcolor=blue,filecolor=black]{hyperref}
\usepackage[colorinlistoftodos, color=green!40, prependcaption]{todonotes}
\usepackage{ulem}
\usepackage{amsthm}
\usepackage{mathtools}
\usepackage{physics}
\usepackage{graphicx}
\usepackage{adjustbox}
\usepackage{placeins}
\usepackage{lipsum}
\usepackage{csquotes}
\usepackage{blindtext}

\newcolumntype{"}{@{\hskip\tabcolsep\vrule width 1pt\hskip\tabcolsep}}

\newcommand{\beq}{\begin{equation}}
\newcommand{\eeq}{\end{equation}}
\renewcommand{\emph}{\textit}

\usepackage{soul}
\usepackage{graphicx}
\usepackage{dcolumn}
\usepackage{bm}

\usepackage[utf8]{inputenc}
\usepackage[T1]{fontenc}
\usepackage{mathptmx}

\begin{document}

\preprint{AIP/123-QED}

\title[]{Quantum randomness generation via orbital angular momentum modes crosstalk in a ring-core fiber}

\author{Mujtaba Zahidy$^{\, 1}$}
\author{Hamid Tebyanian$^{\, 2}$}
\author{Daniele Cozzolino$^{\, 1}$}
\author{Yaoxin Liu$^{\, 1}$} 
\author{Yunhong Ding$^{\, 1}$}
\author{Toshio Morioka$^{\, 1}$} 
\author{Leif K. Oxenløwe$^{\, 1}$} 
\author{Davide Bacco$^{\, 1}$}
\email{dabac@fotonik.dtu.dk}

\affiliation{$^{1}$ Center for Silicon Photonics for Optical Communications (SPOC), Department of Photonics Engineering, Technical University of Denmark, Kgs. Lyngby, Denmark \\
$^{2}$ Department of Mathematics, University of York, Heslington, York, YO10 5DD, United Kingdom
}

\begin{abstract}\noindent
Genuine random numbers can be produced beyond a shadow of doubt through the intrinsic randomness provided by quantum mechanics theory. While many degrees of freedom have been investigated for randomness generation, not adequate attention has been paid to the orbital angular momentum of light. In this work, we present a quantum random number generator based on the intrinsic randomness inherited from the superposition of orbital angular momentum modes caused by the crosstalk inside a ring-core fiber. We studied two possible cases: a first one, device-dependent, where the system is trusted, and a second one, semi-device-independent, where the adversary can control the measurements. We experimentally realized the former, extracted randomness, and, after privacy amplification, we achieved a generation rate higher than 10 Mbit/s. In addition, we presented a possible realization of the semi-device-independent protocol, using a newly introduced integrated silicon photonic chip. Our work can be considered as a starting point for novel investigations of quantum random number generators based on the orbital angular momentum of light.
\end{abstract}
\maketitle

\section{Introduction}
Quantum Mechanics has provided us with unique resources, many of which were not accessible through classical means. Secure communication empowered by quantum key distribution \cite{BENNETT20147,Bacco2021UpConversionQKD,DaLioExp170}, quantum computing \cite{quantum_comp}, quantum conference key agreement \cite{QCKA}, and quantum electronic voting \cite{QEV} are examples of emerging technologies thanks to quantum mechanics. 
Randomness is the critical ingredient of every secure communication protocol; besides, it has a wide range of applications in science and technology, e.g., simulation and gambling \cite{QRNG_rev}.
A random number generator (RNG) should, in general, be secure and practical, otherwise stated easy-to-implement, affordable, and high-rate. Classical RNGs, also known as pseudo-RNGs (PRNG), are quite practical, however the security of the generated random numbers can be compromised, as the randomness generation is based on deterministic phenomena that are predictable \cite{PRNG_gen}. Thus PRNG cannot meet the high-security needs of highly confidential applications \cite{Ma2016}.
Conversely, quantum mechanics can provide genuine and unpredictable randomness based on its intrinsic probabilistic nature, thus allowing for the realization of a quantum RNG.
Quantum random number generators (QRNG)'s protocols are classified into three main categories: device-dependent (DD) \cite{nu_DD,colbeck2011quantum,IDQ_QRNG}, device-independent (DI) \cite{DI_roger,DI_new,DI_new_weak}, and semi-DI \cite{semi-DI_new,avesani2020,semi-DI_rusca}. In the first category, namely DD protocols, the user trusts the performance of the generator and its experimental apparatus. QRNGs based on these kind of protocols could be very practical and relatively secure compared to the classical PRNGs \cite{QRNG_DD}. On the other hand, in the DI scenarios, randomness can be generated needless of trusting the devices' performances, which also implies the violation of a Bell-type inequality to validate the protocols. Even though DI QRNGs offer the highest level of security, they are very slow and complicated, making them less practical \cite{Ma2016}. Semi-DI protocols are an intermediate approach where, depending on the user's demands, some assumptions can be set on the devices. It should be noted that almost all of the commercial QRNGs are based on DD scenarios as they offer a good level of security with an uncomplicated high-rate device \cite{DD_22,DD_33,DD_44,DD_QRNG_1}.
Different QRNGs can be devised on different degrees of freedom either in a discrete variable (DV) or continuous variable (CV) context with each having its benefits and disadvantages \cite{tebyanian2020semidevice}.

\begin{figure}[h]
\includegraphics[width=\linewidth]{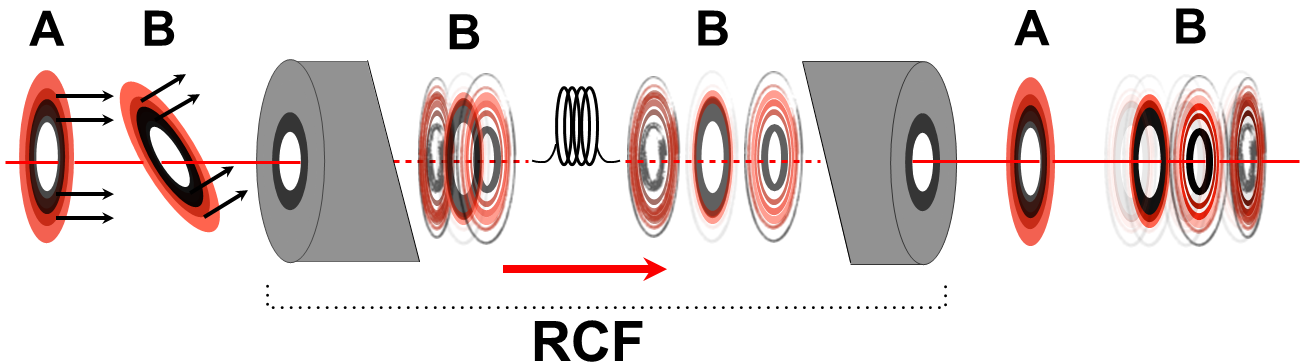}
\caption{A superposition of OAM modes is generated by entering the RCF with a tilted input. Above, input mode `A' will not experience distortion as its wavefront is aligned to the RCF, however, input mode `B' is transformed to a superposition of different modes due to the coupling angle. Various modes are separated along the fiber due to their different group velocities.}
\label{Fig::super}
\centering
\end{figure}

In this paper, we investigate the photons' spatial degree of freedom as a possible source of randomness \cite{Shen2019}. In particular, we propose a QRNG based on the intrinsic crosstalk between orbital angular momentum (OAM) modes in a ring-core fiber (RCF) \cite{RamachandranKristensen_2013_RCF,zahidy2021photonic}, see Figure \ref{Fig::super}.
Such crosstalk is caused by input modes whose wavefront is inclined with respect to the fiber coupler. Indeed, in an RCF, a certain input field would excite several guided modes, depending on its incident angle and field distribution, which can be determined by the overlap of the input field and fiber modes.
Similarly, when a quantum state is coupled into the RCF with an angle, a state in superposition of modes is generated. Measuring such states lead to a probabilistic outcome which is theoretically unpredictable. We leverage on this characteristic to present two QRNG implementations, following a DD and a semi-DI approach. The DD protocol is experimentally realized, and the extracted random numbers are tested for randomness. An experimental proposal for the studied semi-DI protocol is also given in the discussion. 
Our experiment enjoys simplicity and guarantees relatively high generation rate.
In addition, it can be operated either as a no-input generator or as a randomness-expansion implementation.

\section{Protocol}
\label{sec::protocol}
\subsection{Model}
Crosstalk, which diligently is tried to be avoided in many applications \cite{gregg2015conservation,zahidy2021photonic}, can be considered as an intrinsic source of randomness. As simple as an ordinary beam-splitter, crosstalk happening within an RCF can be exploited to extract randomness as it transforms an OAM input state $|M\rangle$ into a superposition of various OAM modes denoted by $|N_j\rangle$,
\begin{equation}
    |M\rangle = \sum_j \lambda_j |N_j\rangle,
\end{equation}
where $|\lambda_j|^2$ is the probability of finding the output state in OAM mode $|N_j\rangle$.
The superposition of OAM modes generated separates in time while propagating through the RCF due to a difference in group velocity for each mode. This mode-to-time mapping allow us to experimentally distinguish each OAM mode with a time of flight measurement from which, subsequently, is possible to extract randomness.

Crosstalk can be exploited to generate randomness both using a DD or semi-DI approach. In the former, which is experimentally realized in this study, a single OAM mode is prepared and coupled to the fiber, while in the latter scenario, an input state is selected from a set of possible modes, which amounts to an increase in randomness and privacy. In what follows, we explain each approach separately. 

\begin{figure}[h]
\includegraphics[width=\linewidth]{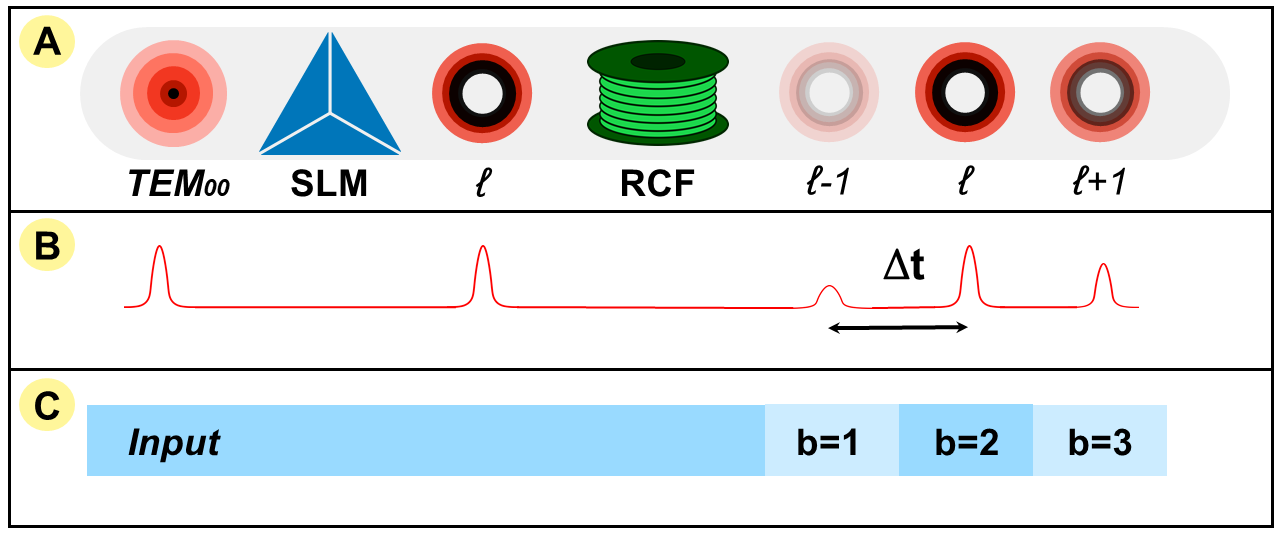}
\caption{\textbf{A)} General model of the experiment: a spatial light modulator (SLM) converts a Gaussian beam to a particular OAM mode. Next, the superposition of OAM modes is generated in the RCF, and the modes are delayed depending on their OAM number. \textbf{B)} Mode-to-time mapping: 
the delay between the OAM modes allows us to distinguish them with a characterized time of flight measurement. Each of the pulses corresponding to an OAM mode is characterized by a specific amplitude. All of them, instead, are separated in time for $T = \Delta t$. \textbf{C)} The random bit, $b$, is extracted based on the time-bin at which the detection events occurred.}
\label{Fig::gen}
\centering
\end{figure}

\subsubsection{Case I: Single Input QRNG} \label{subsec::mode_1}
The simplest QRNG that can be devised is by using a single input inducing a crosstalk such that multiple output modes are excited with probabilities close to $1/d$, with $d$ being the number of modes excited in the RCF.

While this model has simplicity in the preparation stage, the protocol is fully trusted (DD scenario). Yet, we consider the devices' imperfections, e.g. optical devices loss, as the classical side-information known by the adversary. Denoting the set of output modes by $\{N_j\}$ with $|\{N_j\}|=d$, the probability of each outcome in its corresponding time-bin, $b_j$, reads  
\begin{equation}
    P(b_j) = P(N_j|M) \eta_{det} (1-\epsilon)^{d-1},
    \label{EQ::Prob_Mode_1}
\end{equation}
where $\eta_{det}$ is the detector efficiency and $\epsilon$ is the probability of detection error in other time-bins due to dark counts or noise. In low mean photon-number regime, equation (\ref{EQ::Prob_Mode_1}) can be expressed as:
\begin{equation}
    P(b_j) = (1-\xi_j - \xi_j \epsilon) \eta_{det} (1-\epsilon)^{d-1} ,
    \label{EQ::Prob_Mode_1_extended}
\end{equation}
where $\xi_j=|\langle 0 | \alpha_j \rangle|^2=e^{|\alpha_j|^2}$ is the probability of at least 1 photon in the signal and $|\alpha_j|^2=\mu_j$ is the mean photon number at each mode. Ideally, if the initial superposition is balanced we have $\alpha_j = \alpha$ for all $j$.

\subsubsection{Case II: seeded QRNG} \label{subsec::mode_2}
The security of the DD protocol, presented in \ref{subsec::mode_1}, can be improved to a seeded QRNG that increases the privacy of final random numbers generated.
To each input mode, there corresponds a different crosstalk profile,
hence, varying the input mode will enlarge the expected outcome space.

Similar to \ref{subsec::mode_1}, denoting the set of OAM input states $\{M_i\}$ and corresponding outputs $\{N^i_j\}$ with probabilities $\{P(N_j^i|M_i)\}$ and $\{N\}$ be the set of all the outcomes regardless of input mode, one can define the probability of any outcome as:
\begin{equation}
    P(b_j|M_i) = \sum_{i} P(M_i) P(N_j|M_i) \eta_{det} (1-\epsilon)^{d-1},
\end{equation}
where $P(M_i)$ is the probability of selecting the OAM input mode $M_i$ according to a classical input, seed.
\begin{figure}[h]
\includegraphics[width=\linewidth]{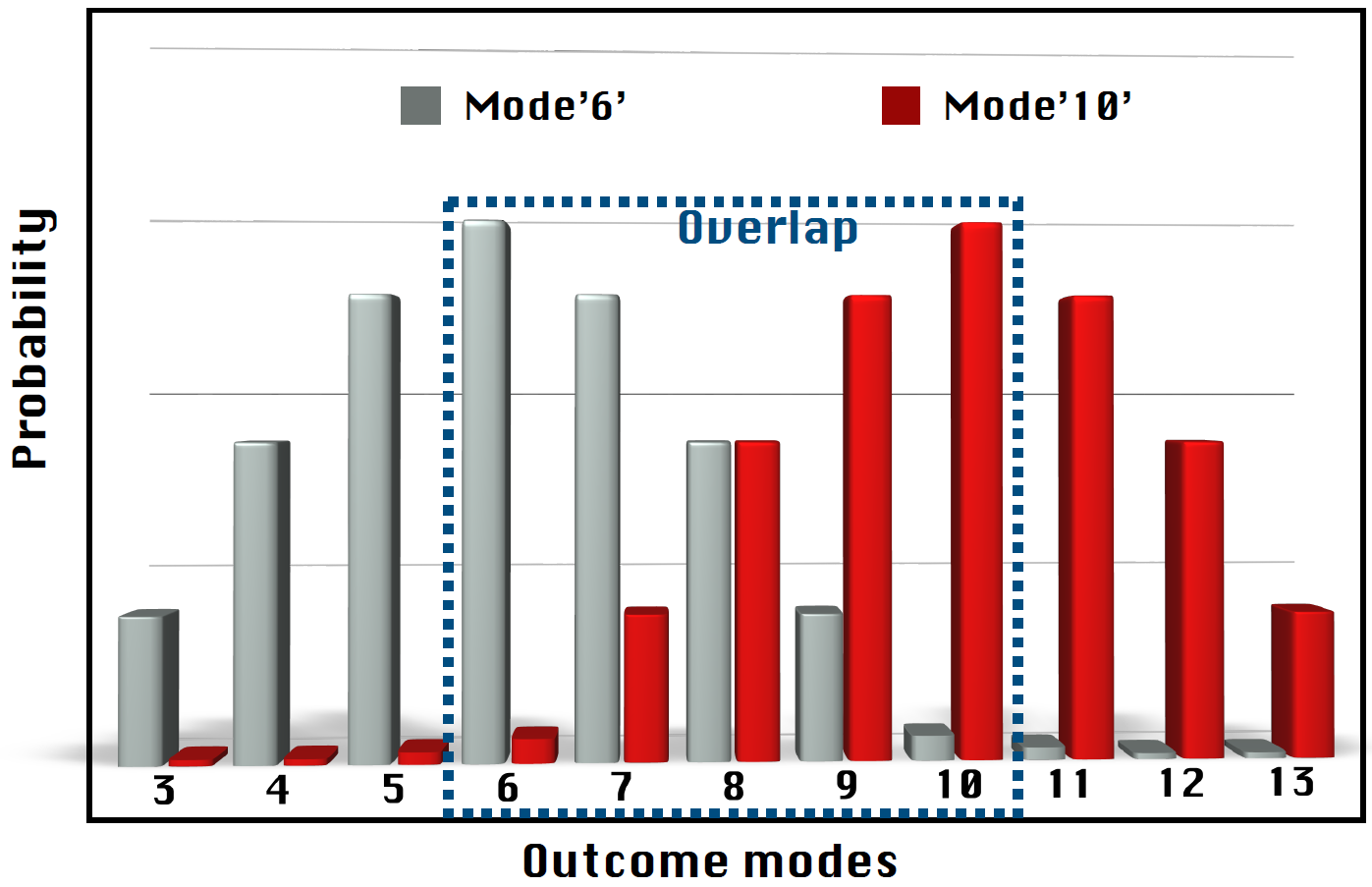}
\caption{Schematic representation of Case II with binary inputs (OAM mode `6' and OAM mode `10'). The gray and red bars show the probability of a click triggered by binary input 0 (mode `6') or 1 (mode `10'), respectively. The dotted area is where the outcomes overlap significantly and the input mode cannot be distinguished unambiguously by the detection events.}
\label{Fig::MDI}
\centering
\end{figure}
Defining multiple inputs provides the ground for introducing a security circumstance in which randomness can be generated without trusting the measurement apparatus, provided the preparation section is trusted and the measurement outcome meets the criteria. 
In general, distinguishing two (or more) neighboring OAM input modes from the measurement result comes with ambiguity. Indeed, the superposition created by each of the input modes has overlaps with the rest, thus one cannot uniquely determine the input mode from the measurement results. 
This uncertainty in the detection limits the adversary's (Eve) power to tamper with the measurement device's output. Eve's presence is detectable in post-processing through a mismatch of the sent and received clicks statistics.
Figure \ref{Fig::MDI} represents a simple example of Case II, with binary input. A click registered in modes $6, 8, 9, 10$, the dotted rectangle, reveals no information about the input OAM mode, as it could be the result of either mode $6$ or $10$.

\begin{figure*}[ht!]
\includegraphics[width=\textwidth]{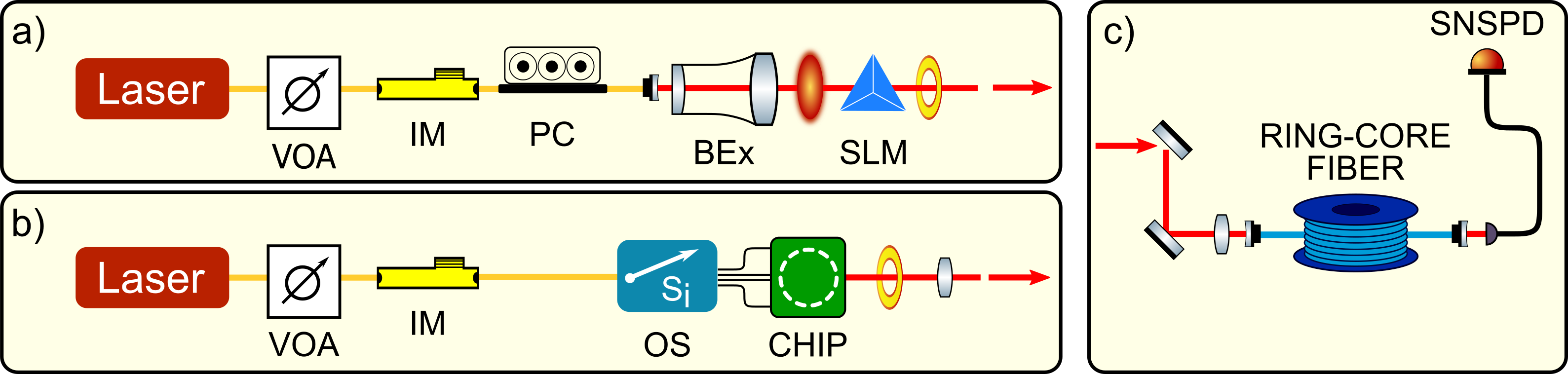}
\caption{\textbf{a)} Experimental setup. VOA: variable optical attenuator; IM: intensity modulator; PC: polarization controller; BEx: beam expander; SLM: spatial light modulator. \textbf{b)} Proposal to implement Case II, \ref{subsec::mode_2}. Chip: integrated silicon photonic chip able to excite OAM modes in the ring-core fiber; OS: optical switch. \textbf{c)} SNSPD: superconducting nano-wire single photon detector.}
\label{Fig::Setup}
\centering
\end{figure*}

\subsection{Security Estimation}
Although an intuitive comprehension of the randomness concept exists, there are alternative ways of defining and understanding it. Fundamentally, the components of a random string should be uniformly distributed, and its elements necessitate to be modeled independently of one another. Otherwise stated, it should be unpredictable; this unpredictability or uncertainty in information can be formulated by information theoretic entropy in a mathematical concept. The most-known method for measuring the informativity of a random variable is defined by Shannon entropy \cite{entropy_aa,shannon2001mathematical}. On the other hand, min-entropy provides a tighter measure to quantify randomness and is the most conservative means of estimating the unpredictability of a set of outcomes \cite{min_entropy_op} and is defined as
\beq
H_{min}(X)=-\log_2[P_{guess}(X)],
\label{EQ::min_ent}
\eeq
where the guessing probability ($ P_{guess}$), defined on the set of possible outcomes, is the maximum probability that an eavesdropper can correctly guess the output of an RNG. Since the worst-case scenario is considered in the min-entropy calculation, it is a more reliable estimate of a system’s randomness.

In this protocol, we assume superposition of OAM modes in the RCF is a non-deterministic phenomenon imposed by the quantum theory.
As shown in Figure \ref{Fig::gen} (B), we consider multiple temporal windows of width $\Delta t$, each corresponding to the arrival time of an OAM mode. Depending on the detected time-bin, the measurement device outputs $b \in \{1,2,3, \dots\}$, e.g., in the case of Figure \ref{Fig::gen} (B), the measurement device returns $b=1$.
Therefore, the measurement output ($b$) is resulting from a probabilistic random phenomenon and forms our probability distribution.

However, the string $b$ is partially deterministic due to classical noises and losses stemming from the experimental apparatus' imperfections. Therefore, to ensure that the extracted random numbers have a quantum origin rather than a classical one, we should exclude all the possible noises introduced by devices imperfections. To account for all these imperfections, conditional min-entropy \cite{conditional_min} as a measure to estimate the amount of extractable randomness in the presence of side-information is employed \cite{conditional_2}. Note that, since the single-input model is the protocol implemented in this article, we assume full control over the device, and no quantum side-information is present. Hence, we limit ourselves to classical side-information, such as laser power fluctuation or losses in the preparation and detection. In this case, the conditional min-entropy on the variable $b$ conditioned on classical side-information ($E$) reads \cite{conditional_3}
\beq
H_{min}(b|E)=-\log_2 P_{guess}(b|E),
\eeq
where 
\beq
P_{guess}(b|E)= \sum_{e}^{} P_{E} (e)\; \max[P(b|E=e)].
\eeq
The above maximization problem can be optimized numerically. The classical side-information probability is experimentally measured by characterizing the experimental devices.
Note that these side-information can only be known to the eavesdropper and can not be controlled or manipulated by her.

\section{Experiment}
The QRNG protocol discussed in section \ref{sec::protocol} is experimentally implemented following Case I, (\ref{subsec::mode_1}) and leveraging on the OAM degree of freedom. 
Photons owning an OAM different from zero are characterized by a helical phase factor $e^{i\ell\varphi}$, where $\varphi$ is the azimuthal angle and $\ell$ is an unbounded integer value representing the quanta of OAM each photon possesses \cite{Allen1992}. Different values of $\ell$ represent different discrete states on which superposition states can be devised.
The experimental setup consists of a continuous laser at 1550 nm, which is carved to form train of pulses at a repetition rate of 12.5 MHz with approximately 2 ns width, see Figure \ref{Fig::Setup} (A). Two cascaded intensity modulator, shown as one in Figure \ref{Fig::Setup}, guarantee high extinction ratio. The electrical signal is generated by an arbitrary waveform generator (AWG) that also provided the signal for clock synchronization. 

The pulses are then collimated and further expanded with a beam expander and modulated by means of a spatial light modulator (SLM) to a definite OAM mode, $\ell=-5$. The resulting signal is then coupled into the ring-core fiber which is capable of carrying up to 12 different OAM modes \cite{RamachandranKristensen_2013_RCF}. The fiber crosstalk stems from two main sources, misalignment of the mode to the RCF and bends and twists along the fiber, with the former being the most dominant factor. The intended superposition is created by exploiting the mode misalignment at the RCF facet. With the help of 2 adjustable mirrors, misalignment is introduced such that an expected crosstalk is observed at the output. For the purpose of this experiment, we achieved a 4-mode crosstalk state where the probability distribution in multiple trials of the experiment is presented in Figure \ref{Fig::ER_Bar}.
Different OAM modes exhibit different group velocities in a medium. Hence, it is possible to distinguish them by a time-of-flight measurement and furthermore, estimate the crosstalk. This estimation can be achieved if the RCF length is long enough to give a noticeable time delay between different modes. If such a condition is satisfied, it is straightforward to allocate different outcomes to distinctive detection time-bin that should be synced with the prepared states a priori.
For this experiment, we used an 800 meter long RCF which amounts to 10 ns temporal separation of adjacent OAM modes. Provided that the optical pulses are short enough, such fiber length gives enough separation to uniquely distinguish the modes.
At the output of the RCF, the pulses are then collimated and coupled into a standard single-mode fiber where they are redirected to a superconducting nano-wire single photon detector (SNSPD) with 83\% detection efficiency and $\approx 50$ dark counts per second. A time to digital converter (TDC) registers the detection events with temporal resolution of 1 ps. 

\section{Result}
\label{sec::result}

\begin{figure}[h]
\includegraphics[width=\linewidth]{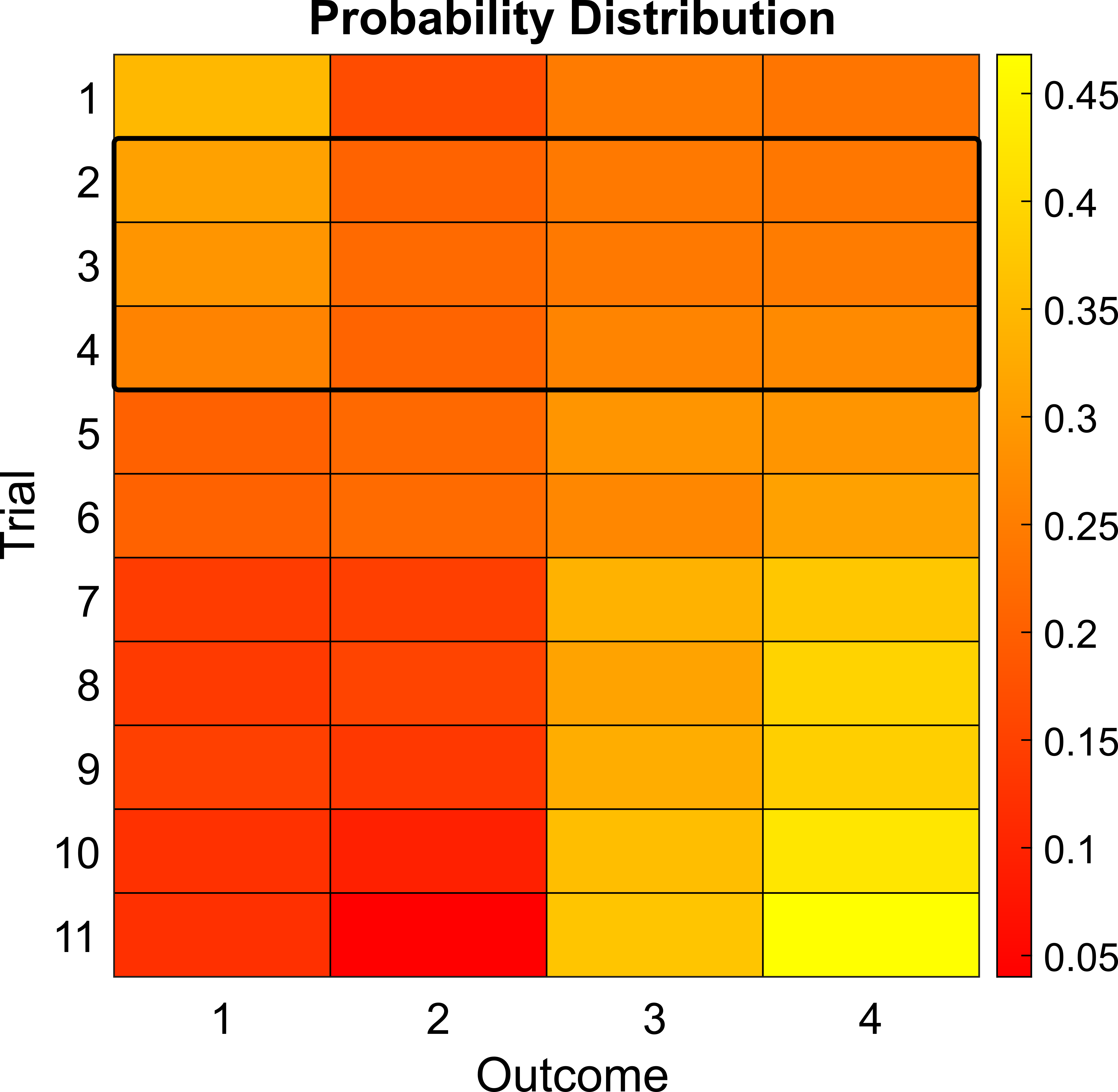}
\caption{Probability distribution of the 4 outcomes obtained in 11 trials. Trials 2 to 4 (see the rectangle) were taken close to the balanced outputs condition, and they give almost uniform distribution.}
\label{Fig::ER_Bar}
\centering
\end{figure}

In this section, we present the results of a test carried out following the Case 1 (\ref{subsec::mode_1}), and performed to certify the protocol as a source of randomness. The Gaussian mode is first converted to OAM $\ell=-5$ and then coupled to the RCF while misalignment is introduced to create the intended crosstalk. We excite 4 dominant modes with close probabilities and several minor modes. Achieving crosstalk to higher number of modes is possible, however, the coupling loss will also increase. In case the input power is limited, this can lead to lower detection rate and a reduction of the rate of the random number generator. The total probability of finding the photon in the 4 dominant modes is more than 98\% in each case, rendering the other excited modes impractical for randomness generation. The 4 dominant excited modes have comparable extinction ratios as also shown by the probability distribution presented in Figure \ref{Fig::ER_Bar}.

We further applied a privacy amplification stage through a Toeplitz randomness extractor with proper parameter chosen based on $H_{min}(b|E)$ \cite{RENNER_thesis,Tomamichel_leftover} and security parameter $\epsilon=10^{-200}$ to remove any non-uniformities. To form the Toeplitz-hashing extractor, PRNG generated by a computer is used. 
With this stage applied to the random numbers, we achieved a generation rate of 10.5 MBit/sec.
It is worth noting that the test performed here was not aimed to exploit the full potential of the proposal but a proof of principle and demonstration. 

\begin{figure}[h]
\includegraphics[width=\linewidth]{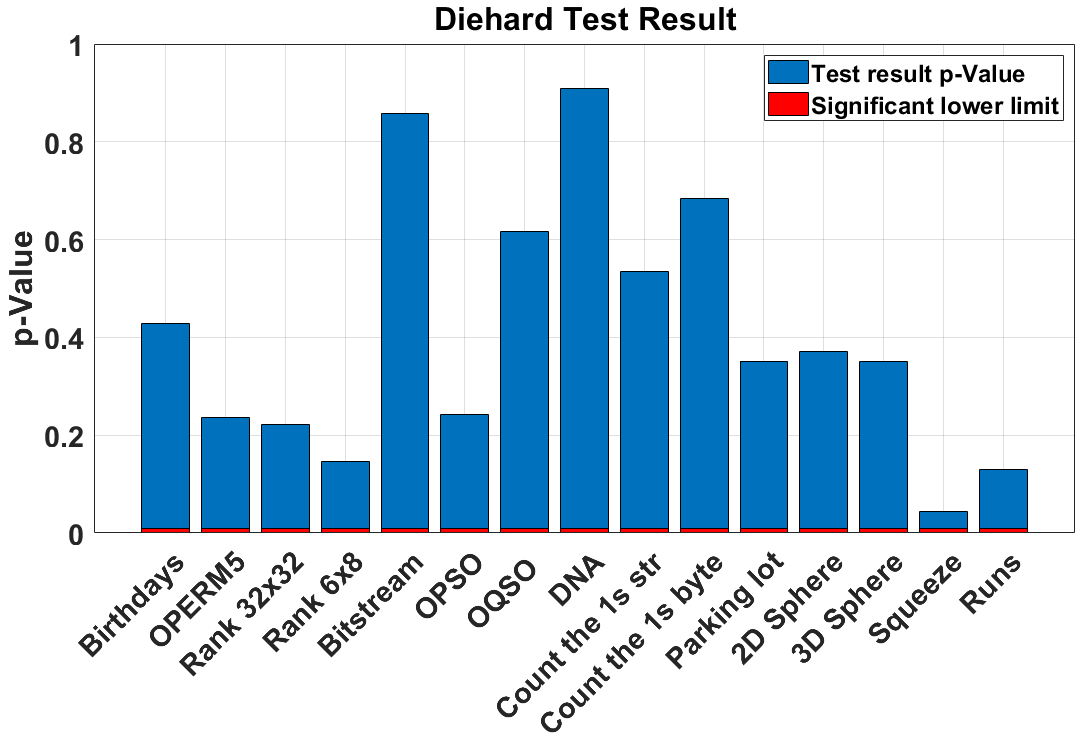}
\caption{Statistical tests, Diehard, performed on the extracted random bits with their respective p-Values. All the tests were passed successfully as the p-Values obtained were higher than the lower limit 0.01. In the picture, only 15 test results are reported.}
\label{Fig::pval}
\centering
\end{figure}

Finally, we performed a set of conventional statistical checks from Diehard to certify the randomness of extracted bits.As shown in Figure \ref{Fig::pval}, the extracted random numbers passed successfully all the Diehard tests executed as the p-Values obtained are higher than the lower threshold 0.01. As such, the possibility to extract quantum randomness from OAM modes crosstalk in a RCF has been demonstrated.

\section{Discussion}
In this work, we successfully shown the possibility of extracting random numbers from crosstalk in an OAM carrying fiber, where the crosstalk is responsible for creating a superposition of OAM modes. 
In particular, after giving the description of the DD and semi-DI protocol, we realized the single input protocol (DD) and we demonstrated the possibility of extracting randomness from crosstalk profiles with a rate of more than 10 Mbit/sec. The quality of the randomness has been certified by executing conventional Diehard tests which have been successfully passed. Nonetheless, it must be noted that passing Diehard tests is a good sign for a genuine quantum randomness extraction, however, they must me regarded as preliminary tests. Indeed, they constitute a necessary but not sufficient step to claim the randomness as truly quantum.

The second and more promising scheme of a QRNG based on OAM mode crosstalk in a RCF has been discussed in Section \ref{subsec::mode_2}. It implies a multi-input source seeding the RCF, thus increasing the overall secuirity (semi-DI case). An experimental proposal of this protocol is shown in Figure \ref{Fig::Setup} (B). It can be realized with a fast switch consisting of intensity modulators and an integrated silicon chip \cite{YaoxinLiuECOC21,zahidy2021photonic}, which enables us to excite different modes at a high rate.

The RCF length is a critical parameter for being able to implement the proposed QRNG schemes. However, we emphasize that having a short enough optical pulse will allow us to reduce the length of the fiber as well as to distinguish more OAM modes.
Indeed, besides those used in our experiment, the RCF can support more modes \cite{gregg2015conservation}, and by exploiting them a significant gain in randomness extraction per detection can be obtained, leading to a more efficient QRNG.

As future plans, along with the experimental realization of the semi-DI protocol, a second approached to both Case I and II can be followed by exploiting superposition states as input to the fiber.
Indeed, using a superposition of multiple OAM modes as the input state of the RCF reduces the overall loss as the fiber will act only as a mean to separate the modes in time, whereas in the proposed idea, the superposition is created at the interface of RCF by misalignments.

We believe this work can be the opening point of a new generation of QRNGs based on OAM, thus creating a wide room for improvements in this promising field.

\vspace{24pt}
\textbf{Acknowledgment:} We thank Roger Colbeck for helpful discussion.

\vspace{10pt}
\textbf{Research funding:} This work is supported by the Center of Excellence SPOC - Silicon Photonics for Optical Communications (ref DNRF123), by the EraNET Cofund Initiatives QuantERA within the European Union’s Horizon 2020 research and innovation program grant agreement No. 731473 (project SQUARE), and by VILLUM FONDEN, QUANPIC (ref. 00025298). H. T. acknowledges the Innovate UK Industrial Strategy Challenge Fund (ISCF), project 106374-49229 AQuRand (Assurance of Quantum Random Number Generators).

\vspace{10pt}
\textbf{Conflict of interest statement:} The authors declare no conflicts of interest regarding this article.

\vspace{10pt}
\textbf{Data availability:}
The data that support the findings of this study are available from the corresponding author upon reasonable request.

\bibliographystyle{IEEEtran}
\bibliography{main}

\end{document}